\documentclass[twocolumn,showpacs,preprintnumbers]{revtex4}%
\usepackage{amssymb}
\usepackage{amsmath}
\usepackage{graphicx}
\usepackage{dcolumn}
\usepackage{bm}
\usepackage{xcolor}
\usepackage{ifthen}
\usepackage[normalem]{ulem}
\usepackage{amsfonts}%
\UseRawInputEncoding
\providecommand{\U}[1]{\protect\rule{.1in}{.1in}}
\providecommand{\U}[1]{\protect\rule{.1in}{.1in}}
\def\showal{1}
\newcommand{\al}[1]{\ifthenelse{\showal=1}{\textcolor{orange}{[[#1]]}}{}}

\newcommand{\eb}[1]{\ifthenelse{\showal=1}{\textcolor{cyan}{[[#1]]}}{}}
\begin{document}
\title{Instability in the Hartmann--Hahn double resonance}
\author{Roei Levi, Sergei Masis and Eyal Buks}
\affiliation{Andrew and Erna Viterbi Department of Electrical Engineering, Technion, Haifa
32000, Israel}
\date{\today }

\begin{abstract}
The Hartmann-Hahn technique allows sensitivity enhancement of magnetic
resonance imaging and spectroscopy by coupling the spins under study to
another spin species that is externally driven. Here we theoretically study
the coupled spins' dynamics, and find that for a certain region of driving
parameters the system becomes unstable. The required conditions for making
this region of instability becoming experimentally accessible are discussed.

\end{abstract}
\maketitle





\section{Introduction}

The technique of cross-polarization (CP) \cite{Pines_569} is widely employed
in magnetic resonance imaging for sensitivity enhancement. Significant CP can
be achieved by applying the so-called Hartmann--Hahn double resonance (HHDR)
\cite{Hartmann1962}. Near the HHDR magnetization can be efficiently
transferred between different spin species \cite{Slichter_Principles}.
Commonly, CP is implemented to enhance the detection sensitivity of a given
spin species under study by applying external driving to another ancilla spin
species having higher polarization. When the Rabi frequency of the ancilla
spins matches the Larmor frequency of the spins under study the so-called
Hartmann--Hahn (HH) matching condition is satisfied \cite{Yang_1} [note that
this is not the same as the matching condition given by Eq. (4) of Ref.
\cite{Hartmann1962}]. In that region a significant CP can be obtained. In
thermal equilibrium the initial polarization of the ancilla spins is
determined by their gyromagnetic ratio and the temperature
\cite{Abragam_1441,Abragam_Principles}. The initial polarization can be
further enhanced when the technique of optically--induced spin polarization
(OISP) can be applied \cite{London2013}.

Here we theoretically study back-reaction effects near the HHDR. A stability
analysis is performed by a linearization of the coupled Bloch equations for
the two spins, one of which is externally-driven. We calculate a correction to
the effective damping rate of the undriven spin, which is induced by the
coupling to the driven one. Analytical results are validated against numerical
calculations. A region of instability, inside which the two-spin system is
expected to exhibit self-excited oscillation (SEO), is identified. The
experimental feasibility of reaching this instability region is discussed.
Related effects of Sisyphus cooling, amplification, lasing and SEO have been
theoretically predicted in other systems having a similar retarded response
\cite{Glenn_195454,Grajcar_612,DeVoogd_42239,Ella_1210_6902,Ramos_193602,Wang_053853}%
.

\section{Dipolar back reaction}

Consider two two-level systems (TLS) having a mutual coupling that is
characterized by a coupling coefficient $g$. The first TLS, which is labelled
as '$\mathrm{a}$', has a relatively low angular frequency $\omega
_{\mathrm{a}0}$ in comparison with the angular frequency $\omega_{\mathrm{b}%
0}$ of the second TLS, which is labelled as '$\mathrm{b}$', and which is
externally driven. It is assumed that the state of the system can be
characterized by the vector of coordinates $\bar{P}=\left(  P_{1},P_{2}%
,P_{3},P_{4},P_{5},P_{6}\right)  ^{\mathrm{T}}$, where $\left(  P_{1}%
,P_{2},P_{3}\right)  =\left(  P_{\mathrm{a}+},P_{\mathrm{a}-},P_{\mathrm{a}%
z}\right)  $ and $\left(  P_{4},P_{5},P_{6}\right)  =\left(  P_{\mathrm{b}%
+},P_{\mathrm{b}-},P_{\mathrm{b}z}\right)  $ are the Bloch vectors of the
first and second TLS, respectively. It is further assumed that the vector of
coordinates $\bar{P}$ satisfies a set of coupled Bloch equations
\cite{Solomon1955}, which are expressed as%
\begin{equation}
\frac{\mathrm{d}P_{n}}{\mathrm{d}t}+\Theta_{n}\left(  \bar{P}\right)
=F_{n}\;, \label{eom P}%
\end{equation}
where $n\in\left\{  1,2,\cdots,6\right\}  $, the functions $\Theta_{n}\left(
\bar{P}\right)  $ (to be specified later) are time independent (which is
possible provided that a rotating frame is used for the driven TLS) and
$F_{n}$ represent fluctuating noise terms having vanishing average values. Let
$\bar{P}_{0}$ be a fixed point, for which $\Theta_{n}\left(  \bar{P}%
_{0}\right)  =0$ for all $n\in\left\{  1,2,\cdots,6\right\}  $. Fluctuations
around the fixed point are governed by%
\begin{equation}
\frac{\mathrm{d}\bar{P}^{\prime}}{\mathrm{d}t}+J\bar{P}^{\prime}=\bar{F}\;,
\label{eom p}%
\end{equation}
where the vector of relative coordinates $\bar{P}^{\prime}=\left(
P_{1}^{\prime},P_{2}^{\prime},P_{3}^{\prime},P_{4}^{\prime},P_{5}^{\prime
},P_{6}^{\prime}\right)  ^{\mathrm{T}}$ is defined by $\bar{P}^{\prime}%
=\bar{P}-\bar{P}_{0}$, the vector of noise terms $\bar{F}$ is given by
$\bar{F}=\left(  F_{1},F_{2},F_{3},F_{4},F_{5},F_{6}\right)  ^{\mathrm{T}}$
and the $6\times6$ Jacobian matrix $J$ at the fixed point $\bar{P}_{0}$ is
defined by $J_{\mathrm{m,n}}=\partial\Theta_{\mathrm{m}}/\partial P_{n}$. The
Jacobian matrix $J$ at the fixed point is expressed in a block form as%
\begin{equation}
J=\left(
\begin{tabular}
[c]{|l|c|}\hline
$J_{\mathrm{x}}$ & $gV_{\mathrm{xy}}$\\\hline
\multicolumn{1}{|c|}{$gV_{\mathrm{yx}}$} & \multicolumn{1}{|l|}{$J_{\mathrm{y}%
}$}\\\hline
\end{tabular}
\ \right)  \;. \label{Jacobian}%
\end{equation}
The subspace corresponding to the subscript $\mathrm{x}$ ($\mathrm{y}$) is
henceforth referred to as the system (ancilla) subspace. For convenience, the
system subspace is chosen to be of dimension $2$ (corresponding to the
transverse variables $P_{\mathrm{a}+}$ and $P_{\mathrm{a}-}$ of spin
'$\mathrm{a}$') and the ancilla subspace of dimension $4$ (corresponding to
the variables $P_{\mathrm{a}z}$, $P_{\mathrm{b}+}$, $P_{\mathrm{b}-}$ and
$P_{\mathrm{b}z}$).

Applying the Fourier transform to Eq. (\ref{eom p}) yields (Fourier angular
frequency is denoted by $\omega$ and lower case $f$ and $p$ denote the Fourier
transform of uppercase $F$ and $P$ variables, respectively)%
\begin{equation}
\left(
\begin{tabular}
[c]{|l|c|}\hline
$J_{\mathrm{x}}-i\omega$ & $gV_{\mathrm{xy}}$\\\hline
\multicolumn{1}{|c|}{$gV_{\mathrm{yx}}$} & \multicolumn{1}{|l|}{$J_{\mathrm{y}%
}-i\omega$}\\\hline
\end{tabular}
\right)  \left(
\begin{array}
[c]{c}%
\bar{p}_{\mathrm{x}}\left(  \omega\right) \\
\bar{p}_{\mathrm{y}}\left(  \omega\right)
\end{array}
\right)  =\left(
\begin{array}
[c]{c}%
\bar{f}_{\mathrm{x}}\left(  \omega\right) \\
\bar{f}_{\mathrm{y}}\left(  \omega\right)
\end{array}
\right)  \;, \label{FT eom p_a p_b}%
\end{equation}
where
\begin{align*}
\bar{p}_{\mathrm{x}}\left(  \omega\right)   &  =\left(  p_{1}\left(
\omega\right)  ,p_{1}^{\ast}\left(  -\omega\right)  \right)  ^{\mathrm{T}%
}\;,\\
\bar{p}_{\mathrm{y}}\left(  \omega\right)   &  =\left(  p_{3}\left(
\omega\right)  ,p_{4}\left(  \omega\right)  ,p_{4}^{\ast}\left(
-\omega\right)  ,p_{6}\left(  \omega\right)  \right)  ^{\mathrm{T}}\;,
\end{align*}
and%
\begin{align*}
\bar{f}_{\mathrm{x}}\left(  \omega\right)   &  =\left(  f_{1}\left(
\omega\right)  ,f_{1}^{\ast}\left(  -\omega\right)  \right)  ^{\mathrm{T}%
}\;,\\
\bar{f}_{\mathrm{y}}\left(  \omega\right)   &  =\left(  f_{3}\left(
\omega\right)  ,f_{4}\left(  \omega\right)  ,f_{4}^{\ast}\left(
-\omega\right)  ,f_{6}\left(  \omega\right)  \right)  ^{\mathrm{T}}\;,
\end{align*}
[note that the $n=2$ ($n=5$) equation of (\ref{eom P}) is the complex
conjugate of the $n=1$ ($n=4$) equation of (\ref{eom P})]. Multiplying the
first [second] row of blocks of Eq. (\ref{FT eom p_a p_b}) by $\chi
_{\mathrm{x}}\left(  \omega\right)  \equiv\left(  J_{\mathrm{x}}%
-i\omega\right)  ^{-1}$ [$\chi_{\mathrm{y}}\left(  \omega\right)
\equiv\left(  J_{\mathrm{y}}-i\omega\right)  ^{-1}$] yields $\bar
{p}_{\mathrm{x}}\left(  \omega\right)  +g\chi_{\mathrm{x}}\left(
\omega\right)  V_{\mathrm{xy}}\bar{p}_{\mathrm{y}}\left(  \omega\right)
=\chi_{\mathrm{x}}\left(  \omega\right)  \bar{f}_{\mathrm{x}}\left(
\omega\right)  $ and $\bar{p}_{\mathrm{y}}\left(  \omega\right)
+g\chi_{\mathrm{y}}\left(  \omega\right)  V_{\mathrm{yx}}\bar{p}_{\mathrm{x}%
}\left(  \omega\right)  =\chi_{\mathrm{y}}\left(  \omega\right)  \bar
{f}_{\mathrm{y}}\left(  \omega\right)  $, and thus $\bar{p}_{\mathrm{x}%
}\left(  \omega\right)  $ can be expressed as $\bar{p}_{\mathrm{x}}\left(
\omega\right)  =\chi_{\mathrm{x},\mathrm{eff}}\left(  \omega\right)  \bar
{f}_{\mathrm{x}}\left(  \omega\right)  $, where $\chi_{\mathrm{x}%
,\mathrm{eff}}\left(  \omega\right)  $ is given by [the term proportional to
$\bar{f}_{\mathrm{y}}\left(  \omega\right)  $ is disregarded, since it does
not affect the expectation value of $\chi_{\mathrm{x},\mathrm{eff}}\left(
\omega\right)  $]%
\begin{align}
\chi_{\mathrm{x},\mathrm{eff}}\left(  \omega\right)   &  =\left(  1-g^{2}%
\chi_{\mathrm{x}}\left(  \omega\right)  V_{\mathrm{xy}}\chi_{\mathrm{y}%
}\left(  \omega\right)  V_{\mathrm{yx}}\right)  ^{-1}\chi_{\mathrm{x}}\left(
\omega\right) \nonumber\\
&  =\left(  J_{\mathrm{x}}-i\omega-g^{2}V_{\mathrm{xy}}\chi_{\mathrm{y}%
}\left(  \omega\right)  V_{\mathrm{yx}}\right)  ^{-1}\;. \label{chi_a,eff}%
\end{align}

The expression for $\chi_{\mathrm{x},\mathrm{eff}}\left(  \omega\right)  $
given by Eq. (\ref{chi_a,eff}) suggests that the coupling effectively shifts
the (complex) resonance frequency of the undriven spin (i.e. the first TLS
labeled as '$\mathrm{a}$'). To lowest nonvanishing order in the coupling
coefficient $g$, the underlying mechanism responsible for this shift, as
revealed by Eq. (\ref{chi_a,eff}), is a three-step feedback process. In the
first step, consider the case where spin '$\mathrm{a}$' undergoes precession
with small amplitude at its own Larmor (i.e. resonance) frequency. The term
$gV_{\mathrm{yx}}$ in Eq. (\ref{chi_a,eff}) represents the driving applied to
the ancilla system due to the precession of spin '$\mathrm{a}$', and the term
$\chi_{\mathrm{y}}\left(  \omega\right)  $ in Eq. (\ref{chi_a,eff}) represents
the corresponding response of the ancilla to this driving (the second step).
This response of the ancilla gives rise to a feedback driving applied to spin
'$\mathrm{a}$' occurring in the third step, where the feedback coupling is
represented by the term $gV_{\mathrm{xy}}$ in Eq. (\ref{chi_a,eff}).

The coupling-induced feedback driving applied to spin '$\mathrm{a}$' can be
expressed as a sum of two orthogonal quadratures, both oscillating at the
Larmor frequency of spin '$\mathrm{a}$'. The first one is in-phase with the
precession of spin '$\mathrm{a}$', and the second one, which occurs due to
retardation in the response of the ancilla to the precession of spin
'$\mathrm{a}$', is out of phase. The in-phase quadratures gives rise to a
change in the real part of the effective resonance frequency of the undriven
spin '$\mathrm{a}$' (i.e. a change in its effective Larmor frequency), whereas
the change in the effective damping rate is proportional to the amplitude of
the out of phase quadrature.

In general, dipolar interaction is represented by a Hamiltonian $\mathcal{H}%
_{\mathrm{d}}$ containing terms proportional to operators having the form
$S_{\mathrm{a},i}S_{\mathrm{b},j}$, where $S_{\mathrm{a},i}$ ($S_{\mathrm{b}%
,j}$) is the $i$'th ($j$'th) component of spin '$\mathrm{a}$' ('$\mathrm{b}$')
angular momentum operator $\mathbf{S}_{\mathrm{a}}$ ($\mathbf{S}_{\mathrm{b}}%
$) (see \cite{Slichter_Principles} p. 66). Terms proportional to the
longitudinal component of $\mathbf{S}_{\mathrm{a}}$ are disregarded since they
do not contribute to effective driving at the Larmor frequency of spin
'$\mathrm{a}$'. Moreover, terms proportional to transverse components of
$\mathbf{S}_{\mathrm{b}}$ can be disregarded as well, provided that the Larmor
frequency of spin '$\mathrm{b}$' is much higher than the Larmor frequency of
spin '$\mathrm{a}$'. In this limit the driving applied to spin '$\mathrm{b}$'
due to the precession of spin '$\mathrm{a}$' can be considered as slow, and
consequently its transverse component has a weak effect compared to the effect
of its longitudinal component (which effectively modulates the Larmor
frequency of spin '$\mathrm{b}$'). When only dominant terms are kept the
dipolar coupling Hamiltonian $\mathcal{H}_{\mathrm{d}}$ becomes $\mathcal{H}%
_{\mathrm{d}}=2g\hbar^{-1}\left(  S_{\mathrm{a+}}+S_{\mathrm{a-}}\right)
S_{\mathrm{bz}}$.

The Hamiltonian $\mathcal{H}$ of the closed system is given by%
\begin{equation}
\mathcal{H}=\omega_{\mathrm{a}0}S_{\mathrm{az}}+\omega_{\mathrm{b}%
0}S_{\mathrm{bz}}+\omega_{\mathrm{b}1}\left(  S_{\mathrm{b+}}+S_{\mathrm{b-}%
}\right)  +\mathcal{H}_{\mathrm{d}}\;, \label{Hamiltonian}%
\end{equation}
where the driving amplitude and angular frequency are denoted by
$\omega_{\mathrm{b1}}$ and $\omega_{\mathrm{p}}=\omega_{\mathrm{b0}}%
+\Delta_{\mathrm{b}}$, respectively ($\Delta_{\mathrm{b}}$ is the driving
detuning), the operators $S_{\mathrm{a\pm}}$ are given by $S_{\mathrm{a\pm}%
}=S_{\mathrm{ax}}\pm iS_{\mathrm{ay}}$, and the rotated operators
$S_{\mathrm{b\pm}}$ are given by $S_{\mathrm{b\pm}}=\left(  S_{\mathrm{bx}}\pm
iS_{\mathrm{by}}\right)  e^{\mp i\omega_{\mathrm{p}}t}$. The Heisenberg
equation of motion $\mathrm{d}O/\mathrm{d}t=-i\hbar^{-1}\left[  O,\mathcal{H}%
\right]  +\partial O/\partial t$, where $O$ is a given observable, together
with the spin commutation relations $\left[  S_{z},S_{\pm}\right]  =\pm\hbar
S_{\pm}$ and $\left[  S_{+},S_{-}\right]  =2\hbar S_{z}$ yield (overdot
denotes a time derivative) $\dot{S}_{\mathrm{a}+}-i\omega_{\mathrm{a}%
0}S_{\mathrm{a}+}+4ig\hbar^{-1}S_{\mathrm{az}}S_{\mathrm{bz}}=0$, $\dot
{S}_{\mathrm{az}}+2ig\hbar^{-1}\left(  S_{\mathrm{a}+}-S_{\mathrm{a}-}\right)
S_{\mathrm{bz}}=0$, $\dot{S}_{\mathrm{b}+}+i\left(  \Delta_{\mathrm{b}%
}-2g\hbar^{-1}\left(  S_{\mathrm{a}+}+S_{\mathrm{a}-}\right)  \right)
S_{\mathrm{b}+}+2i\omega_{\mathrm{b}1}S_{\mathrm{bz}}=0$ and $\dot
{S}_{\mathrm{bz}}+i\omega_{\mathrm{b}1}\left(  S_{\mathrm{b}+}-S_{\mathrm{b}%
-}\right)  =0$.

The interaction with the environment is accounted for by assuming that the
closed system is weakly coupled to thermal baths at thermal equilibrium. The
coupling turns the deterministic equations of motion for the spin operators
into Langevin equations\ containing both damping and fluctuating terms. By
applying thermal averaging, which is denoted by $\left\langle {}\right\rangle
$, a set of coupled equations can be derived.

The coupling terms (i.e. the terms proportional to $g$) in the above-derived
evolution equations for the operators $S_{\mathrm{a}+}$, $S_{\mathrm{az}}$ and
$S_{\mathrm{b}+}$ have the form $gAB$, where $A$ ($B$) is an operator of spin
'$\mathrm{a}$' ('$\mathrm{b}$'). The following holds $\left\langle
AB\right\rangle =\left\langle A\right\rangle \left\langle B\right\rangle
+\left\langle V_{\mathrm{AB}}\right\rangle $, where $V_{\mathrm{AB}}=\left(
A-\left\langle A\right\rangle \right)  \left(  B-\left\langle B\right\rangle
\right)  $. In the mean field approximation the term $\left\langle
V_{\mathrm{AB}}\right\rangle $ is disregarded. Note that the following holds
$g\left\langle AB\right\rangle =g\left\langle A_{0}\right\rangle \left\langle
B_{0}\right\rangle +O\left(  g^{2}\right)  $, where $A_{0}$ ($B_{0}$)
represents the operator $A$ ($B$) in the decoupling limit of $g\rightarrow0$,
hence the mean field approximation is consistent with our assumption that the
coupling coefficient $g$ is small. This approximation greatly
simplifies the problem, since it allows the description of the dynamics in
terms of the vector $\bar{P}=\left(  P_{\mathrm{a}+},P_{\mathrm{a}%
-},P_{\mathrm{az}},P_{\mathrm{b}+},P_{\mathrm{b}-},P_{\mathrm{bz}}\right)
^{\mathrm{T}}$, where $\left(  \hbar/2\right)  P_{\mathrm{az}}=\left\langle
S_{\mathrm{az}}\right\rangle $, $\left(  \hbar/2\right)  P_{\mathrm{bz}%
}=\left\langle S_{\mathrm{bz}}\right\rangle $, $\hbar P_{\mathrm{a}\pm
}=\left\langle S_{\mathrm{a}\pm}\right\rangle $ and $\hbar P_{\mathrm{b}\pm
}=\left\langle S_{\mathrm{b}\pm}\right\rangle $. The vector $\bar{P}$ is
determined by $2$ real numbers $P_{\mathrm{a}z}$ and $P_{\mathrm{b}z}$, and
$2$ complex numbers $P_{\mathrm{a}+}=P_{\mathrm{a}-}^{\ast}$ and
$P_{\mathrm{b}+}=P_{\mathrm{b}-}^{\ast}$, whereas more variables are needed
for treating the general case (the density operator of a two-spin system is
determined by $15$ real numbers).

Note that for product states, for which $\left\langle V_{\mathrm{AB}%
}\right\rangle =0$, the mean field approximation becomes exact. It can be used
provided that the lifetime of entangled states is much shorter than all single
spin lifetimes. When this assumption cannot be made a more general analysis,
which does not exclude entanglement, is needed. The so-called
concurrence \cite{Wootters_2245} allows quantifying the entanglement. An
expression for a critical temperature $T_{\mathrm{c}}$, above which in steady
state the spin-spin system becomes separable (i.e. entanglement vanishes) has
been derived in \cite{Guiroga_032308,Sinaysky_062301}. Near the HH matching
condition the critical temperature is approximately given by $T_{\mathrm{c}%
}\simeq\hbar g/k_{\mathrm{B}}$, where $k_{\mathrm{B}}$ is the Boltzmann
constant. Hence, for the vast majority of magnetic resonance experimental
setups, for which the temperature $T\gg T_{\mathrm{c}}$ (recall that $g$
represents dipolar coupling), entanglement can be safely disregarded.

In the mean field approximation the functions $\overline{\Theta}\left(
\bar{P}\right)  =\left(  \Theta_{\mathrm{1}},\Theta_{\mathrm{2}}%
,\Theta_{\mathrm{3}},\Theta_{\mathrm{4}},\Theta_{\mathrm{5}},\Theta
_{\mathrm{6}}\right)  ^{\mathrm{T}}$ are given by $\Theta_{\mathrm{1}}%
=\Theta_{\mathrm{2}}^{\ast}=\left(  \gamma_{\mathrm{a2}}-i\omega_{\mathrm{a0}%
}\right)  P_{\mathrm{a+}}+igP_{\mathrm{az}}P_{\mathrm{bz}}$, $\Theta
_{\mathrm{3}}=\gamma_{\mathrm{a1}}\left(  P_{\mathrm{az}}-P_{\mathrm{az}%
,\mathrm{s}}\right)  +2ig\left(  P_{\mathrm{a+}}-P_{\mathrm{a-}}\right)
P_{\mathrm{bz}}$, $\Theta_{\mathrm{4}}=\Theta_{\mathrm{5}}^{\ast}=\left(
\gamma_{\mathrm{b2}}+i\Delta_{\mathrm{b}}\right)  P_{\mathrm{b+}}%
+i\omega_{\mathrm{b1}}P_{\mathrm{bz}}-2ig\left(  P_{\mathrm{a+}}%
+P_{\mathrm{a-}}\right)  P_{\mathrm{b+}}$ and $\Theta_{\mathrm{6}}%
=\gamma_{\mathrm{b1}}\left(  P_{\mathrm{bz}}-P_{\mathrm{bz},\mathrm{s}%
}\right)  +2i\omega_{\mathrm{b1}}\left(  P_{\mathrm{b+}}-P_{\mathrm{b-}%
}\right)  $, where $\gamma_{\mathrm{a1}}$ ($\gamma_{\mathrm{b1}}$) is the
longitudinal relaxation rate of the undriven (driven) spin,\ $\gamma
_{\mathrm{a2}}$ ($\gamma_{\mathrm{b2}}$) is the transverse relaxation rate of
the undriven (driven) spin, $P_{\mathrm{az},\mathrm{s}}=-\tanh\left(
\hbar\omega_{\mathrm{a}0}/2k_{\mathrm{B}}T\right)  $ ($P_{\mathrm{bz}%
,\mathrm{s}}=-\tanh\left(  \hbar\omega_{\mathrm{b}0}/2k_{\mathrm{B}}T\right)
$)\ is the value of $P_{\mathrm{az}}$ ($P_{\mathrm{b}z}$) in thermal
equilibrium, and $k_{\mathrm{B}}T$ is the thermal energy. Note that in steady
state the expectation values of the transverse components of the undriven spin
'$\mathrm{a}$' vanish (for the decoupled case $g=0$), i.e. $P_{\mathrm{a}%
+},=P_{\mathrm{a}-}=0$.

The derivation below is mainly devoted to the analytical inversion of the
matrix $J_{\mathrm{y}}-i\omega$, which, in turn, allows the evaluation of
$\chi_{\mathrm{x},\mathrm{eff}}\left(  \omega\right)  $ according to Eq.
(\ref{chi_a,eff}). The matrices $J_{\mathrm{x}}~$($2\times2$) and
$J_{\mathrm{y}}$ ($4\times4$) are given by%
\begin{equation}
J_{\mathrm{x}}=\left(
\begin{array}
[c]{cc}%
\gamma_{\mathrm{a2}}-i\omega_{\mathrm{a}0} & 0\\
0 & \gamma_{\mathrm{a2}}+i\omega_{\mathrm{a}0}%
\end{array}
\right)  \;, \label{J_a=}%
\end{equation}%
\begin{equation}
J_{\mathrm{y}}=\left(
\begin{array}
[c]{cccc}%
\gamma_{\mathrm{a1}} & 0 & 0 & 0\\
0 & \gamma_{\mathrm{b2}}+i\Delta_{\mathrm{b}} & 0 & i\omega_{\mathrm{b1}}\\
0 & 0 & \gamma_{\mathrm{b2}}-i\Delta_{\mathrm{b}} & -i\omega_{\mathrm{b1}}\\
0 & 2i\omega_{\mathrm{b1}} & -2i\omega_{\mathrm{b1}} & \gamma_{\mathrm{b1}}%
\end{array}
\right)  ~, \label{J_b=}%
\end{equation}
and the coupling matrices $V_{\mathrm{xy}}~$($2\times4$) and $V_{\mathrm{yx}%
}~$($4\times2$) are given by%
\begin{equation}
V_{\mathrm{xy}}=\left(
\begin{array}
[c]{cccc}%
iP_{\mathrm{bz}} & 0 & 0 & iP_{\mathrm{az}}\\
-iP_{\mathrm{bz}} & 0 & 0 & -iP_{\mathrm{az}}%
\end{array}
\right)  \;, \label{V_ab pmz}%
\end{equation}%
\begin{equation}
V_{\mathrm{yx}}=\left(
\begin{array}
[c]{cc}%
2iP_{\mathrm{bz}} & -2iP_{\mathrm{bz}}\\
-2iP_{\mathrm{b+}} & -2iP_{\mathrm{b+}}\\
2iP_{\mathrm{b-}} & 2iP_{\mathrm{b-}}\\
0 & 0
\end{array}
\right)  \;. \label{V_ba pmz}%
\end{equation}
To lowest nonvanishing order in $g$ the blocks $V_{\mathrm{xy}}$ and
$V_{\mathrm{yx}}$ are evaluated by replacing all variables $\bar{P}$ by their
averaged steady-state values in the absence of coupling, which are denoted as
$\bar{P}_{0}=\left(  P_{\mathrm{a}+0},P_{\mathrm{a}-0},P_{\mathrm{a}%
z0},P_{\mathrm{b}+0},P_{\mathrm{b}-0},P_{\mathrm{b}z0}\right)  $. These
averaged steady-state values are evaluated in appendix \ref{appFPBE}, and it
is found that [see Eq. (\ref{SSS B})]%
\begin{equation}
P_{\mathrm{b}z0}=\frac{\left(  1+\frac{\Delta_{\mathrm{b}}^{2}}{\gamma
_{\mathrm{b}2}^{2}}\right)  P_{\mathrm{b}z,\mathrm{s}}}{1+\frac{4\omega
_{\mathrm{b}1}^{2}}{\gamma_{\mathrm{b}1}\gamma_{\mathrm{b}2}}+\frac
{\Delta_{\mathrm{b}}^{2}}{\gamma_{\mathrm{b}2}^{2}}}\;, \label{P_bz0}%
\end{equation}%
\begin{equation}
P_{\mathrm{b}+0}=\frac{\frac{\omega_{\mathrm{b}1}}{\gamma_{\mathrm{b}2}%
}\left(  -\frac{\Delta_{\mathrm{b}}}{\gamma_{\mathrm{b}2}}-i\right)
P_{\mathrm{b}z,\mathrm{s}}}{1+\frac{4\omega_{\mathrm{b}1}^{2}}{\gamma
_{\mathrm{b}1}\gamma_{\mathrm{b}2}}+\frac{\Delta_{\mathrm{b}}^{2}}%
{\gamma_{\mathrm{b}2}^{2}}}\;, \label{P_b+0}%
\end{equation}
$P_{\mathrm{b}+0}^{\ast}=P_{\mathrm{b}-0}$, $P_{\mathrm{a}z0}=P_{\mathrm{a}%
z,\mathrm{s}}$ and $P_{\mathrm{a}+0}=P_{\mathrm{a}-0}=0$ (since the first spin
is not driven).

Next, the effective susceptibility matrix $\chi_{\mathrm{x},\mathrm{eff}%
}\left(  \omega\right)  $ is evaluated at the resonance frequency of the first
TLS $\omega_{\mathrm{a}0}$. The following holds [see Eq. (\ref{J_b=})]%
\begin{gather}
\chi_{\mathrm{y}}\left(  \omega_{\mathrm{a}0}\right)  =\left(  J_{\mathrm{y}%
}-i\omega_{\mathrm{a}0}\right)  ^{-1}\nonumber\\
=\frac{1}{D_{\mathrm{L}}}\left(
\begin{array}
[c]{cccc}%
\frac{D_{\mathrm{L}}}{D_{0}} & 0 & 0 & 0\\
0 & D_{2}D_{3}+2\omega_{\mathrm{b}1}^{2} & 2\omega_{\mathrm{b}1}^{2} &
-i\omega_{\mathrm{b}1}D_{2}\\
0 & 2\omega_{\mathrm{b}1}^{2} & D_{1}D_{3}+2\omega_{\mathrm{b}1}^{2} &
i\omega_{\mathrm{b}1}D_{1}\\
0 & -2i\omega_{\mathrm{b}1}D_{2} & 2i\omega_{\mathrm{b}1}D_{1} & D_{1}D_{2}%
\end{array}
\right)  \ , \label{chi_b}%
\end{gather}
where $D_{0}=\gamma_{\mathrm{a}1}-i\omega_{\mathrm{a}0}$, $D_{1}%
=\gamma_{\mathrm{b2}}+i\Delta_{\mathrm{b}}-i\omega_{\mathrm{a}0}$,
$D_{2}=\gamma_{\mathrm{b2}}-i\Delta_{\mathrm{b}}-i\omega_{\mathrm{a}0}$,
$D_{3}=\gamma_{\mathrm{b}1}-i\omega_{\mathrm{a}0}$ and $D_{\mathrm{L}}%
=D_{1}D_{2}D_{3}+2\omega_{\mathrm{b}1}^{2}\left(  D_{1}+D_{2}\right)  $. As
can be seen from Eq. (\ref{chi_a,eff}), only the diagonal elements of
$V_{\mathrm{xy}}\chi_{\mathrm{y}}\left(  \omega\right)  V_{\mathrm{yx}}$
contribute to the eigenvalues of $\chi_{\mathrm{x},\mathrm{eff}}\left(
\omega\right)  $ to lowest nonvanishing order in $g$ (second order). To the
same order the effective complex frequency of the first TLS is $\omega
_{\mathrm{a}0}+i\gamma_{\mathrm{a}2}-i\Upsilon_{\mathrm{a}}$, where%
\begin{equation}
\Upsilon_{\mathrm{a}}=g^{2}\left(  V_{\mathrm{xy}}\chi_{\mathrm{y}}\left(
\omega\right)  V_{\mathrm{yx}}\right)  _{11}\;. \label{Upsilon_a}%
\end{equation}
Substituting the corresponding coupling matrices $V_{\mathrm{xy}}$ and
$V_{\mathrm{yx}}$ leads to [see Eqs. (\ref{V_ab pmz}), (\ref{V_ba pmz}),
(\ref{chi_b}) and (\ref{P_b+0})]
\begin{equation}
\Upsilon_{\mathrm{a}}=-2g^{2}\left(  \frac{P_{\mathrm{bz0}}^{2}}{D_{0}%
}+2i\omega_{\mathrm{b1}}P_{\mathrm{az0}}\frac{D_{2}P_{\mathrm{b+0}}%
+D_{1}P_{\mathrm{b-0}}}{D_{\mathrm{L}}}\right)  \;. \label{Upsilon_a V1}%
\end{equation}

The determinant $D_{\mathrm{L}}$ can be expressed as $D_{\mathrm{L}}%
/\omega_{\mathrm{a}0}^{3}=\left(  \gamma_{\mathrm{b}1}/\omega_{\mathrm{a}%
0}\right)  \eta$, where $\eta=\eta^{\prime}+i\eta^{\prime\prime}$,
$\eta^{\prime}=\Delta_{\mathrm{b}}^{2}/\omega_{\mathrm{a}0}^{2}-\left(
1+\left(  2\gamma_{\mathrm{b2}}/\gamma_{\mathrm{b}1}\right)  \left(
1-2\omega_{\mathrm{b}1}^{2}/\omega_{\mathrm{a}0}^{2}\right)  -\gamma
_{\mathrm{b2}}^{2}/\omega_{\mathrm{a}0}^{2}\right)  $, $\eta^{\prime\prime
}=\left(  1-\left(  2\gamma_{\mathrm{b}1}/\omega_{\mathrm{a}0}+\gamma
_{\mathrm{b2}}/\omega_{\mathrm{a}0}\right)  \left(  \gamma_{\mathrm{b2}%
}/\omega_{\mathrm{a}0}\right)  -\omega_{\mathrm{R}}^{2}/\omega_{\mathrm{a}%
0}^{2}\right)  /\left(  \gamma_{\mathrm{b}1}/\omega_{\mathrm{a}0}\right)  $,
and $\omega_{\mathrm{R}}$, which is given by%
\begin{equation}
\omega_{\mathrm{R}}=\sqrt{4\omega_{\mathrm{b}1}^{2}+\Delta_{\mathrm{b}}^{2}%
}\;,
\end{equation}
is the Rabi frequency of the driven spins, thus Eq. (\ref{Upsilon_a V1}) can
be rewritten as%
\begin{align}
\frac{\omega_{\mathrm{a}0}\Upsilon_{\mathrm{a}}}{2g^{2}}=  &  \frac
{\frac{4\left(  1+\frac{2i\gamma_{\mathrm{b2}}}{\omega_{\mathrm{a}0}}\right)
\Delta_{\mathrm{b}}\omega_{\mathrm{b1}}^{2}P_{\mathrm{az,s}}P_{\mathrm{b}%
z,\mathrm{s}}}{\gamma_{\mathrm{b}2}^{2}\gamma_{\mathrm{b}1}\eta}}%
{1+\frac{4\omega_{\mathrm{b}1}^{2}}{\gamma_{\mathrm{b}1}\gamma_{\mathrm{b}2}%
}+\frac{\Delta_{\mathrm{b}}^{2}}{\gamma_{\mathrm{b}2}^{2}}}\nonumber\\
+  &  \frac{\left(  1+\frac{\Delta_{\mathrm{b}}^{2}}{\gamma_{\mathrm{b}2}^{2}%
}\right)  \left(  \frac{\gamma_{\mathrm{a}1}}{\omega_{\mathrm{a}0}}-i\right)
P_{\mathrm{b}z,\mathrm{s}}^{2}}{\left(  1+\frac{4\omega_{\mathrm{b}1}^{2}%
}{\gamma_{\mathrm{b}1}\gamma_{\mathrm{b}2}}+\frac{\Delta_{\mathrm{b}}^{2}%
}{\gamma_{\mathrm{b}2}^{2}}\right)  ^{2}\left(  1+\frac{\gamma_{\mathrm{a1}%
}^{2}}{\omega_{\mathrm{a0}}^{2}}\right)  }\;.\nonumber\\
&  \label{Upsilon_a=}%
\end{align}
The effective damping rate of spin '$\mathrm{a}$' is given by $\gamma
_{\mathrm{a2}}\left(  1+\alpha_{\mathrm{a}}\right)  $, where $\alpha
_{\mathrm{a}}=-\operatorname{Re}\left(  \Upsilon_{\mathrm{a}}\right)
/\gamma_{\mathrm{a2}}$. The contribution of the term in the second line of Eq.
(\ref{Upsilon_a=}) to $\alpha_{\mathrm{a}}$ can be disregarded provided that
$\gamma_{\mathrm{a1}}\ll\omega_{\mathrm{a0}}$.

\begin{figure}[ptb]
\begin{center}
\includegraphics[
trim=0.000000in 0.000000in -0.018166in 0.000000in,
height=5.9119in,
width=3.2889in
]{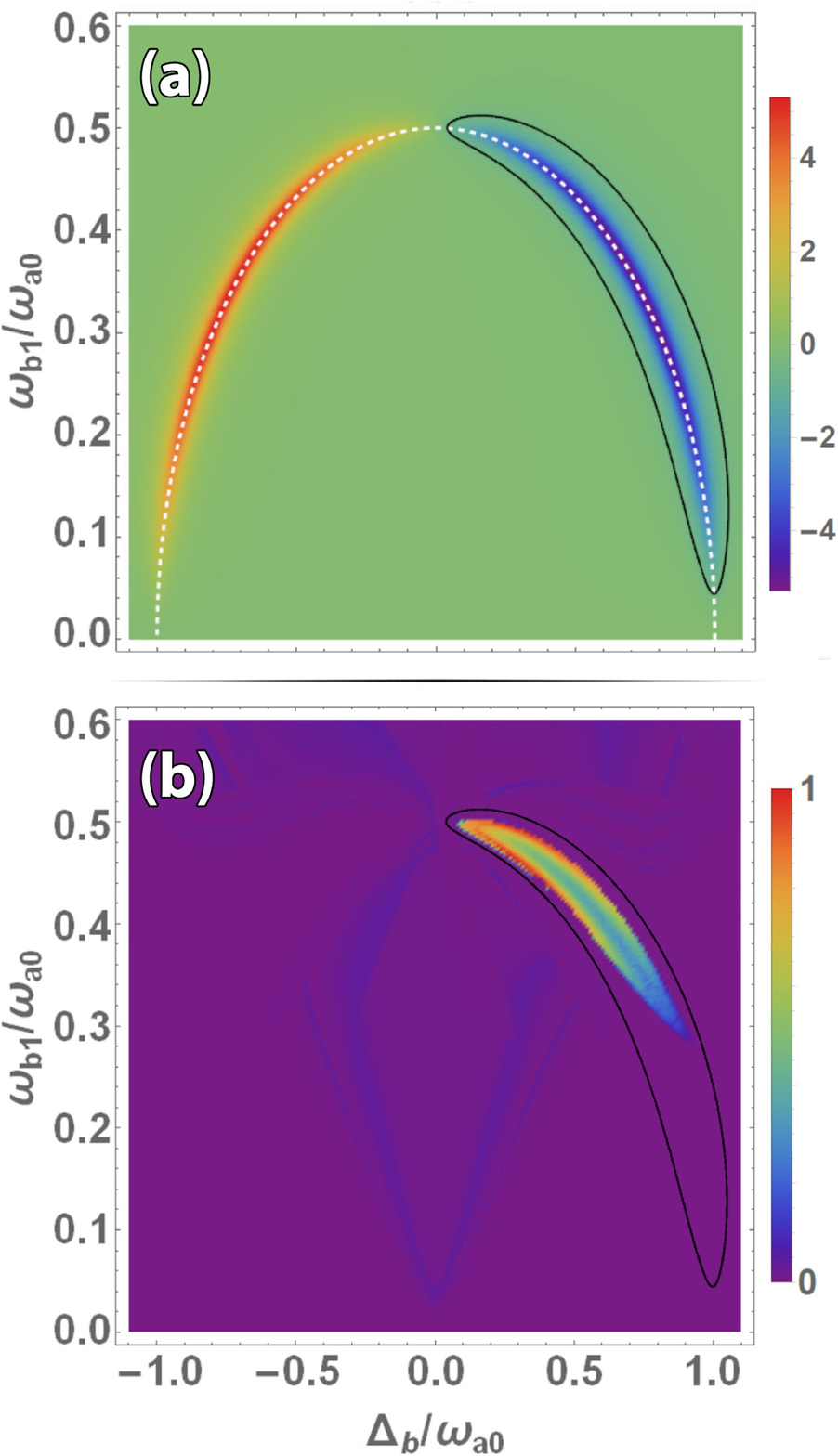}
\end{center}
\caption{The effective damping rate of spin '$\mathrm{a}$'. (a) Color-coded
plot of $\alpha_{\mathrm{a}}$ vs. $\Delta_{\mathrm{b}}/\omega_{\mathrm{a}0}$
and $\omega_{\mathrm{b}1}/\omega_{\mathrm{a}0}$ with parameters $g/\omega
_{\mathrm{a}0}=1.0$, $\gamma_{\mathrm{a1}}/\omega_{\mathrm{a}0}=10^{-2}$,
$\gamma_{\mathrm{a2}}/\omega_{\mathrm{a}0}=10^{-4}$, $\gamma_{\mathrm{b}%
1}/\omega_{\mathrm{a}0}=3.7\times10^{-3}$, $\gamma_{\mathrm{b}2}%
/\omega_{\mathrm{a}0}=3.7\times10^{-2},$ $P_{\mathrm{a}z,\mathrm{s}}%
=-5\times10^{-4}$ and $P_{\mathrm{b}z,\mathrm{s}}=-1$. (b) Numerical solution
for the normalized steady state amplitude of $P_{\mathrm{a+}}$ vs.
$\Delta_{\mathrm{b}}/\omega_{\mathrm{a}0}$ and $\omega_{\mathrm{b}1}%
/\omega_{\mathrm{a}0}$, with the same parameters as in (a). The fluctuating
noise terms $\bar{F}$ are disregarded in the numerical calculation.}%
\label{Fig_P_a_plus_ss}%
\end{figure}

The dependence according to Eq. (\ref{Upsilon_a=}) of $\alpha_{\mathrm{a}}$ on
the normalized detuning $\Delta_{\mathrm{b}}/\omega_{\mathrm{a}0}$ and on the
normalized driving amplitude $\omega_{\mathrm{b}1}/\omega_{\mathrm{a}0}$ is
shown in Fig. \ref{Fig_P_a_plus_ss}(a) (the term proportional to
$P_{\mathrm{bz0}}^{2}$ is disregarded since it is assumed that $\gamma
_{\mathrm{a}1}\ll\omega_{\mathrm{a}0}$). The parameters that have been used
for generating the plot are listed in the figure caption. Spin '$\mathrm{b}$'
is assumed to be fully polarized, for instance by OISP. Also, the curve along
which $\alpha_{\mathrm{a}}=-1$ is shown as a solid black curve on the same
plot. This curve labels the border between the regions of positive and
negative effective damping rates for the undriven spin. When the driving is
red-detuned , i.e. when $\Delta_{\mathrm{b}}$ is negative, the change in
damping rate $\gamma_{\mathrm{a2}}$ is positive, and consequently spin cooling
is expected to occur \cite{Aspelmeyer_1391}. The opposite behavior occurs with
blue detuning, i.e. when $\Delta_{\mathrm{b}}$ is positive. Specifically, SEO
is expected in the area enclosed by the black curve. Along this curve the
system undergoes a Hopf bifurcation \cite{Hassard_Hopf}.

For both red and blue detuning, large change in the effective spin damping
rate occurs near the overlaid dashed white line in Fig. \ref{Fig_P_a_plus_ss},
along which the Rabi frequency $\omega_{\mathrm{R}}$ coincides with
$\omega_{\mathrm{a}0}$, i.e. $\Delta_{\mathrm{b}}=\pm\sqrt{\omega
_{\mathrm{a}0}^{2}-4\omega_{\mathrm{b}1}^{2}}$, and the HH matching condition
is satisfied. This behavior can be explained by noticing that $\left\vert
\eta^{\prime\prime}\right\vert \ll1$ along the dashed line, i.e. when
$\omega_{\mathrm{R}}=\omega_{\mathrm{a}0}$, and consequently $\left\vert
\alpha_{\mathrm{a}}\right\vert $ obtains a peak.

The underlying mechanism responsible for the change in the effective damping
rate of the undriven spin is similar to a related mechanism occurring in
optomechanical cavities \cite{Aspelmeyer_1391}. The change in the effective
damping rate of the system under study (i.e. spin '$\mathrm{a}$') is
attributed to the retardation in the response of the driven ancilla (i.e. spin
'$\mathrm{b}$') to fluctuation in the state of spin '$\mathrm{a}$'. Both
effects of cooling and heating are attributed to imbalance between fluctuation
and dissipation \cite{Aspelmeyer_1391} occurring due to the change in the
effective damping rate of spin '$\mathrm{a}$'.

The analytic result given by Eq. (\ref{Upsilon_a=}) was validated against a
numerical simulation of a time dependent solution of the equations of motion
[see Eq. (\ref{eom P})]. A plot for the normalized steady state amplitude of
$P_{\mathrm{a+}}$ vs. the normalized detuning $\Delta_{\mathrm{b}}%
/\omega_{\mathrm{a}0}$ and the normalized driving amplitude $\omega
_{\mathrm{b}1}/\omega_{\mathrm{a}0}$ is shown in Fig. \ref{Fig_P_a_plus_ss}%
(b). The undriven spin experiences SEO in the region of negative effective
damping rate (encircled by the black curve). Deviation between the region of
SEO that is obtained by the solid black curve and the one that is obtained
from the numerical calculation is attributed to the term $P_{\mathrm{bz0}}%
^{2}/D_{0}$ that was neglected in Eq. (\ref{Upsilon_a=}).

As was mentioned above, our analysis is based on the mean field approximation,
which, in turn, is based on the assumption that entanglement is nearly fully
suppressed. The more general case can be explored using the system's master
equation. In general, the master equation contains terms originating from the
unitary evolution generated by the Hamiltonian $\mathcal{H}$
(\ref{Hamiltonian}) of the closed system, and terms originating from the
interaction with the environment. In some cases only linear terms associated
with the interaction with the environment are kept. For these cases the master
equation can be expressed in the form given by Eq. (\ref{dk/dt=}) of appendix
\ref{appLME}. The matrix $G$ in Eq. (\ref{dk/dt=}) represents linear damping.
In appendix \ref{appLME} we show that for these cases instabilities are
excluded provided that all diagonal matrix elements of $G$ are positive [see
inequality (\ref{kappa' <=})]. This observation suggests that a master
equation having the form given by Eq. (\ref{dk/dt=}) is inapplicable for our case.

Grabert has shown that the invalidity of the quantum regression hypothesis
gives rise to a nonlinear term in the master equation of a general quantum
system \cite{Grabert_161}. This nonlinear term, which is ignored in many
publications, is not included in Eq. (\ref{dk/dt=}). Using general expressions
derived in \cite{Ottinger_052119,Ottinger_10006} a Grabert master equation can
be derived for the two spins problem under study here. Moreover, for this
problem , additional nonlinear terms have to be added to the master equation
\cite{Yukalov_9232,Prataviera_01}, since external driving is applied, and
consequently the transverse coordinates of spin 'b' cannot be assumed to be
small. In the derivation of the master equation it is important to note that
additivity of decay rates may break down for bipartite decoherence
\cite{Yu_140403}. Since the resultant nonlinear master equation cannot be
expressed in the form given by Eq. (\ref{dk/dt=}), instabilities cannot be
generally excluded. Detailed analysis based on the nonlinear master equation
is kept outside the scope of the current paper.

\section{Experimental feasibility}

The experimental feasibility to reach the instability threshold occurring when
$\alpha_{\mathrm{a}}=-1$ is discussed below. Consider the case where the HH
condition $\omega_{\mathrm{R}}=\omega_{\mathrm{a}0}$ is satisfied [see the
dashed white curve in Fig. \ref{Fig_P_a_plus_ss}(a)]. As is demonstrated by
Fig. \ref{Fig_P_a_plus_ss}, the largest change in $\alpha_{\mathrm{a}}$
typically occurs when the detuning $\left\vert \Delta_{\mathrm{b}}\right\vert
$ and and driving amplitude $\omega_{\mathrm{b}1}$ are of the same order of
magnitude (i.e. $\left\vert \Delta_{\mathrm{b}}\right\vert \simeq
\omega_{\mathrm{b}1}\simeq\omega_{\mathrm{a}0}$). When the following holds
$\omega_{\mathrm{a}0}\ll\omega_{\mathrm{b}0}$, $\gamma_{\mathrm{b}1}\ll
\gamma_{\mathrm{b}2}\ll\left\vert \Delta_{\mathrm{b}}\right\vert $ and
$\left\vert \Delta_{\mathrm{b}}\right\vert \simeq\omega_{\mathrm{b}1}%
\simeq\omega_{\mathrm{a}0}$,\ the threshold condition $\alpha_{\mathrm{a}}=-1$
yields the requirement $\kappa P_{\mathrm{a}z,\mathrm{s}}P_{\mathrm{b}%
z,\mathrm{s}}\simeq1$, where $\kappa=g^{2}\gamma_{\mathrm{b}1}/\left(
\gamma_{\mathrm{a2}}\gamma_{\mathrm{b2}}^{2}\right)  $ is the cooperativity
parameter [see Eq. (\ref{Upsilon_a=})].

As an example, consider two nearby defects in a diamond lattice
\cite{Alfasi_214111}. The first one having Larmor angular frequency
$\omega_{\mathrm{a}0}$ is a negatively charged nitrogen vacancy (NV$^{-}$)
defect \cite{Doherty_1}, and the second one is a nitrogen substitutional
defect (P1) \cite{Cook_99} having Larmor angular frequency $\omega
_{\mathrm{b}0}$. An externally applied magnetic field $B$ parallel to the NV
axis can be used to tune both $\omega_{\mathrm{a}0}$ and $\omega_{\mathrm{b}%
0}$. Two spin states belonging to the NV$^{-}$ spin triplet ground state
become nearly degenerate near the magnetic field value of $B=102%
\operatorname{mT}%
$. In that region the angular frequency $\omega_{\mathrm{b}0}$ of the
electronic-like P1 transitions is about $\omega_{\mathrm{b}0}=\gamma
_{\mathrm{e}}B=2\pi\times2.9%
\operatorname{GHz}%
$, where $\gamma_{\mathrm{e}}=2\pi\times28.03%
\operatorname{GHz}%
\operatorname{T}%
^{-1}$ is the electron spin gyromagnetic ratio. For this value of
$\omega_{\mathrm{b}0}$ the P1 electronic spin defects can be nearly fully
polarized, i.e. $\left\vert P_{\mathrm{b}z,\mathrm{s}}\right\vert \simeq1$, by
cooling down the sample well below a temperature of about $0.07%
\operatorname{K}%
$.

In practice, the value of the NV transition frequency $\omega_{\mathrm{a}0}$
(which can be tuned by the externally applied magnetic field) is limited due
to the HH matching condition by the maximum possible value of the driving
amplitude $\omega_{\mathrm{b}1}$ that can be experimentally achieved. For the
case where a high quality factor microwave resonator is employed for driving
the P1 spins \cite{Wang_053853}, a value of about $\omega_{\mathrm{a}0}%
=\omega_{\mathrm{R}}\simeq2\pi\times50%
\operatorname{MHz}%
$, is reachable. This value is too small to allow making $\left\vert
P_{\mathrm{a}z,\mathrm{s}}\right\vert $ becoming of order unity using cooling
only. However the condition $\left\vert P_{\mathrm{a}z,\mathrm{s}}\right\vert
\simeq1$ can be satisfied using the technique of OISP
\cite{Robledo_025013,Redman_3420}.

The dipolar coupling coefficient $g$ between the NV$^{-}$ electron spin and
the P1 electron spin is given by $g/2\pi\simeq3.6%
\operatorname{GHz}%
\left(  r_{\mathrm{d}}/a_{\mathrm{d}}\right)  ^{-3}$
\cite{Slichter_Principles}, where $r_{\mathrm{d}}$ is the NV$^{-}$-P1
distance, $a_{\mathrm{d}}=3.57%
\operatorname{\text{\AA}}%
$ is the lattice constant of diamond, and it is assumed for simplicity that
the angle between the line joining the two defects and the NV axis vanishes.
When both spins are fully polarized and for the typical values of
$\gamma_{\mathrm{a2}}=\gamma_{\mathrm{b}2}=2\pi\times0.1%
\operatorname{MHz}%
$, $\gamma_{\mathrm{b}1}=2\pi\times0.01%
\operatorname{MHz}%
$, the threshold condition $\alpha_{\mathrm{a}}=-1$ is satisfied when
$r_{\mathrm{d}}\simeq8%
\operatorname{nm}%
$.

In the above example the case of dipolar coupling between two electron spins
localized near different lattice sites was considered. This coupling, however,
depends on the distance between sites, and consequently, its study is
difficult using measurements of ensembles containing many spins. Such
Inhomogeneity is avoided for the case where the same lattice site hosts both
spins. For that case spin '$\mathrm{a}$' is assumed to be a nuclear spin and
spin '$\mathrm{b}$' is an electron spin. For example, for the case of P1
defects in diamond, the dipolar coupling between the nitrogen 14 nuclear spin
$S=1$ and the spin of the localized unpaired electron occupying the same
lattice site gives rise to hyperfine splitting on the order of $100%
\operatorname{MHz}%
$ \cite{Wang_053853}. Such a coupling is sufficiently strong to make the
region of SEO experimentally accessible, and the study of this instability can
be performed using measurements of ensembles containing many P1 defects.

\section{Summary}

Our results demonstrate that a significant change in the effective value of
transverse spin relaxation rate can be induced, provided that the HH matching
condition can be satisfied. Red-detuned driving provides a positive
contribution to the relaxation rate, whereas negative contribution can be
obtained by blue-detuned driving. For the former case this effect can be
utilized for cooling down spins, while the later case of blue detuning may
allow inducing SEO. Operating close to the threshold of SEO, i.e. close to the
point where the total effective damping vanishes, may be useful for sensing
applications, since the system is expected to become highly responsive to
external perturbations near the threshold. It is important to emphasize that our analysis is based on the mean field approximation, and therefore our results are inapplicable for the case where entanglement cannot be disregarded.

We thank one of the referees for suggesting to explore the
stability of the system under study using the master equation. This work was supported by the Israel science foundation, the Israel ministry of science and the security research foundation at Technion.

\appendix

\section{Fixed points of Bloch equations}

\label{appFPBE}

The dynamics of the polarization vector $\mathbf{P}=P_{x}\mathbf{\hat{x}%
}+P_{y}\mathbf{\hat{y}}+P_{z}\mathbf{\hat{z}}$, which describes the state of
the spin system, is governed by the Bloch equations \cite{Slichter_Principles}%
\begin{equation}
\frac{\mathrm{d}\mathbf{P}}{\mathrm{d}t}=\mathbf{P}\times\mathbf{\Omega
}+\mathbf{\gamma}\;, \label{Bloch P}%
\end{equation}
where $\mathbf{\Omega}\left(  t\right)  $ is the rotation vector, which is
proportional to the externally applied magnetic field vector (the factor of
proportionality is called the gyromagnetic ratio). The vector%
\begin{equation}
\mathbf{\gamma}=-\gamma_{2}P_{x}\mathbf{\hat{x}}-\gamma_{2}P_{y}%
\mathbf{\hat{y}}-\gamma_{1}\left(  P_{z}-P_{z,\mathrm{s}}\right)
\mathbf{\hat{z}}\
\end{equation}
represents the contribution of damping, where $\gamma_{1}=1/T_{1}$ and
$\gamma_{2}=1/T_{2}$ are the longitudinal and transverse relaxation rates,
respectively, and where $P_{z,\mathrm{s}}$ is the equilibrium steady state polarization.

Consider the case where the rotation vector $\mathbf{\Omega}\left(  t\right)
$ is taken to be given by%
\begin{equation}
\mathbf{\Omega}\left(  t\right)  =2\omega_{1}\left(  \cos\left(  \omega
t\right)  \mathbf{\hat{x}}+\sin\left(  \omega t\right)  \mathbf{\hat{y}%
}\right)  +\omega_{0}\mathbf{\hat{z}}\ , \label{Omega(t)}%
\end{equation}
where $\omega_{1}$, $\omega$ and $\omega_{0}$ are real constants. For this
case Eq. (\ref{Bloch P}) becomes%
\begin{equation}
\frac{\mathrm{d}\mathbf{P}}{\mathrm{d}t}+M_{\mathrm{B}}\mathbf{P}=\left(
\begin{array}
[c]{c}%
0\\
0\\
\gamma_{1}P_{z,\mathrm{s}}%
\end{array}
\right)  \;,
\end{equation}
where%
\begin{equation}
M_{\mathrm{B}}=\left(
\begin{array}
[c]{ccc}%
\gamma_{2} & -\omega_{0} & 2\omega_{1}\sin\left(  \omega t\right) \\
\omega_{0} & \gamma_{2} & -2\omega_{1}\cos\left(  \omega t\right) \\
2\omega_{1}\sin\left(  \omega t\right)  & -2\omega_{1}\cos\left(  \omega
t\right)  & \gamma_{1}%
\end{array}
\right)  \;.
\end{equation}
The variable transformation%
\begin{equation}
\left(
\begin{array}
[c]{c}%
P_{x}\\
P_{y}%
\end{array}
\right)  =\left(
\begin{array}
[c]{cc}%
e^{i\omega t} & e^{-i\omega t}\\
-ie^{i\omega t} & ie^{-i\omega t}%
\end{array}
\right)  \left(
\begin{array}
[c]{c}%
P_{+}\\
P_{-}%
\end{array}
\right)  \;,
\end{equation}
leads to%
\begin{equation}
\frac{\mathrm{d}}{\mathrm{d}t}\left(
\begin{array}
[c]{c}%
P_{+}\\
P_{-}\\
P_{z}%
\end{array}
\right)  +J\left(
\begin{array}
[c]{c}%
P_{+}\\
P_{-}\\
P_{z}%
\end{array}
\right)  =\left(
\begin{array}
[c]{c}%
0\\
0\\
\gamma_{1}P_{z,\mathrm{s}}%
\end{array}
\right)  \;, \label{eom P pm z}%
\end{equation}
where%
\begin{equation}
J=\left(
\begin{array}
[c]{ccc}%
\gamma_{2}-i\Delta & 0 & i\omega_{1}\\
0 & \gamma_{2}+i\Delta & -i\omega_{1}\\
2i\omega_{1} & -2i\omega_{1} & \gamma_{1}%
\end{array}
\right)  \ , \label{J B}%
\end{equation}
and where $\Delta=\omega-\omega_{0}$ is the driving detuning. The steady state
solution of Eq. (\ref{eom P pm z}) is given by%
\begin{align}
\left(
\begin{array}
[c]{c}%
P_{+0}\\
P_{-0}\\
P_{z0}%
\end{array}
\right)   &  =J^{-1}\left(
\begin{array}
[c]{c}%
0\\
0\\
\gamma_{1}P_{z,\mathrm{s}}%
\end{array}
\right) \nonumber\\
&  =\left(
\begin{array}
[c]{c}%
\frac{\frac{\omega_{1}}{\gamma_{2}}\left(  -\frac{\Delta}{\gamma_{2}%
}-i\right)  }{1+\frac{4\omega_{1}^{2}}{\gamma_{1}\gamma_{2}}+\frac{\Delta^{2}%
}{\gamma_{2}^{2}}}\\
\frac{\frac{\omega_{1}}{\gamma_{2}}\left(  -\frac{\Delta}{\gamma_{2}%
}+i\right)  }{1+\frac{4\omega_{1}^{2}}{\gamma_{1}\gamma_{2}}+\frac{\Delta^{2}%
}{\gamma_{2}^{2}}}\\
\frac{1+\frac{\Delta^{2}}{\gamma_{2}^{2}}}{1+\frac{4\omega_{1}^{2}}{\gamma
_{1}\gamma_{2}}+\frac{\Delta^{2}}{\gamma_{2}^{2}}}%
\end{array}
\right)  P_{z,\mathrm{s}}\ . \label{SSS B}%
\end{align}

\section{Linear master equation}

\label{appLME}

Consider a closed quantum system having Hilbert space of dimension
$d_{\mathrm{H}}$, whose Hamiltonian is given by $\mathcal{H}\dot{=}\hbar
\Omega_{\mathrm{H}}$, where the $d_{\mathrm{H}}\times d_{\mathrm{H}}$ matrix
$\Omega_{\mathrm{H}}$ is Hermitian and time independent. It is assumed that
the master equation for the system's reduced density matrix $\rho$ can be
expressed as%
\begin{align}
\frac{\mathrm{d}\rho}{\mathrm{d}t}  &  =i\left[  \rho,\Omega_{\mathrm{H}%
}\right]  -\gamma_{\mathrm{E}}\left[  Q,\left[  Q,\rho\right]  \right]
\nonumber\\
&  -\eta_{\mathrm{E}}\gamma_{\mathrm{E}}\left[  Q,\left[  Q,\Omega
_{\mathrm{H}}\right]  \right]  \;,\nonumber\\
&  \label{LME}%
\end{align}
where the coefficient $\gamma_{\mathrm{E}}>0$ is a damping rate,
$\eta_{\mathrm{E}}>0$ is dimensionless, and the dimensionless Hermitian matrix
$Q$ represents the interaction with the system's environment. In this appendix
it is shown that any master equation having this form is stable provided that
$d_{\mathrm{H}}$ is finite.

The density matrix can be expressed as%
\begin{equation}
\rho=\frac{1}{d_{\mathrm{H}}}+\bar{k}\cdot\bar{\lambda}\;, \label{rho ex}%
\end{equation}
where $\bar{k}=\left(  k_{1},k_{2},\cdots,k_{d_{\mathrm{H}}^{2}-1}\right)  $
and $\bar{\lambda}=\left(  \lambda_{1},\lambda_{2},\cdots,\lambda
_{d_{\mathrm{H}}^{2}-1}\right)  $. The $d_{\mathrm{H}}^{2}-1$ Hermitian and
trace-less $d_{\mathrm{H}}\times d_{\mathrm{H}}$ matrices $\lambda_{n}$, which
span the SU($d_{\mathrm{H}}$) Lie algebra, satisfy the orthogonality relation%
\begin{equation}
\frac{\operatorname{Tr}\left(  \lambda_{a}\lambda_{b}\right)  }{2}=\delta
_{ab}\;. \label{Tr(lambda_a*lambda_b)=}%
\end{equation}
For example, for the case of 2-level (3-level) system, i.e. for $d_{\mathrm{H}%
}=2$ ($d_{\mathrm{H}}=3$), the $d_{\mathrm{H}}^{2}-1=3$ Pauli\ ($d_{\mathrm{H}%
}^{2}-1=8$ Gell-Mann) matrices can be used. Note that the condition
$\operatorname{Tr}\rho^{2}=d_{\mathrm{H}}^{-1}+2\left\vert \bar{k}\right\vert
^{2}\leq1$ implies that $\left\vert \bar{k}\right\vert ^{2}\leq\left(
1-d_{\mathrm{H}}^{-1}\right)  /2$.

With the help of the orthogonality relation (\ref{Tr(lambda_a*lambda_b)=}) and
the general trace identity $\operatorname{Tr}\left(  XY\right)
=\operatorname{Tr}\left(  YS\right)  $ the master equation (\ref{LME}) can be
expressed as (repeated index implies summation)%
\begin{align}
\frac{\mathrm{d}k_{a}}{\mathrm{d}t}  &  =\frac{i}{2}\operatorname{Tr}\left(
\Omega_{\mathrm{H}}\left(  \left[  \lambda_{a},\lambda_{b}\right]  \right)
\right)  k_{b}\nonumber\\
&  -\frac{\gamma_{\mathrm{E}}}{2}\operatorname{Tr}\left(  -\left[
Q,\lambda_{b}\right]  \left[  Q,\lambda_{a}\right]  \right)  k_{b}\nonumber\\
&  -\frac{\eta_{\mathrm{E}}\gamma_{\mathrm{E}}}{2}\operatorname{Tr}\left(
\left[  Q,\left[  Q,\Omega_{\mathrm{H}}\right]  \right]  \lambda_{a}\right)
\;,\nonumber\\
&  \label{dk_a/dt}%
\end{align}
or in a matrix form%
\begin{equation}
\frac{\mathrm{d}\bar{k}}{\mathrm{d}t}=\left(  M-G\right)  \bar{k}+\bar{k}%
_{0}\;. \label{dk/dt=}%
\end{equation}

The $\left(  d_{\mathrm{H}}^{2}-1\right)  \times\left(  d_{\mathrm{H}}%
^{2}-1\right)  $ matrix $M$, which represents the unitary evolution governed
by the Hamiltonian of the closed system $\mathcal{H}$, is given by
$M_{a,b}=\left(  i/2\right)  \operatorname{Tr}\left(  \Omega_{\mathrm{H}%
}\left[  \lambda_{a},\lambda_{b}\right]  \right)  $ [see Eq. (\ref{dk_a/dt})].
Note that the matrix $M$ is \textit{real} and \textit{antisymmetric} (or skew
symmetric), i.e. $M^{\mathrm{T}}=-M$. This implies that $\det\left(
M^{\mathrm{T}}\right)  =\det\left(  -M\right)  =\left(  -1\right)
^{d_{\mathrm{H}}^{2}-1}\det M$, hence $\det M=0$ for odd $d_{\mathrm{H}}%
^{2}-1$. Note also that when interaction with the environment is disregarded,
i.e. when $\gamma_{\mathrm{E}}=0$, the following holds $\left(  \mathrm{d}%
\bar{k}/\mathrm{d}t\right)  \cdot\bar{k}=0$, i.e. for this dissipation-less
case the magnitude $\left\vert \bar{k}\right\vert $ of the vector $\bar{k}%
$\ is a constant of the motion.

The $\left(  d_{\mathrm{H}}^{2}-1\right)  \times\left(  d_{\mathrm{H}}%
^{2}-1\right)  $ matrix $G$, which represents linear damping, is given by
$G_{a,b}=\left(  \gamma_{\mathrm{E}}/2\right)  \operatorname{Tr}\left(
-\left[  Q,\lambda_{b}\right]  \left[  Q,\lambda_{a}\right]  \right)  $ [see
Eq. (\ref{dk_a/dt})]. The elements of the vector $\bar{k}_{0}$ are given by
$\left(  \bar{k}_{0}\right)  _{a}=\left(  -\eta_{\mathrm{E}}\gamma
_{\mathrm{E}}/2\right)  \operatorname{Tr}\left(  \left[  Q,\left[
Q,\Omega_{\mathrm{H}}\right]  \right]  \lambda_{a}\right)  $. Note that all
elements of $G$ and $\bar{k}_{0}$ are \textit{real} (recall that, in general,
$i\left[  A,B\right]  $ is Hermitian provided that both $A$ and $B$ are
Hermitian). Moreover, all \textit{diagonal} elements of $G$ are
\textit{positive} (note that $-\left[  Q,\lambda_{b}\right]  \left[
Q,\lambda_{a}\right]  $ is positive-definite for the case $a=b$).

The solution of Eq. (\ref{dk/dt=}) is given by%
\begin{align}
\bar{k}\left(  t\right)   &  =e^{\left(  M-G\right)  t}\bar{k}\left(
0\right)  +\int_{0}^{t}\mathrm{d}t^{\prime}\;e^{\left(  M-G\right)  \left(
t-t^{\prime}\right)  }\bar{k}_{0}\;.\nonumber\\
&  \label{k(t)}%
\end{align}
The system's stability depends on the set of eigenvalues of the matrix $M-G$,
which is denoted by $\mathcal{S}$. The system is stable provided that
$\operatorname{real}\left(  \xi\right)  <0$ for any $\xi\in\mathcal{S}$. For
that case the steady state solution is given by $-\left(  M-G\right)
^{-1}\bar{k}_{0}$.

Let $v=v^{\prime}+iv^{\prime\prime}$ be an eigenvector of $M-G$ with
eigenvalue $\xi=\xi^{\prime}+i\xi^{\prime\prime}$, where $\xi^{\prime}%
,\xi^{\prime\prime}\in\mathbb{R}$ and $v^{\prime},v^{\prime\prime}%
\in\mathbb{R}^{d_{\mathrm{H}}^{2}-1}$ ($\mathbb{R}$ denotes the set of real
numbers)%
\begin{equation}
\left(  M-G\right)  v=\xi v\;.\label{v kappa}%
\end{equation}
It is shown below that the system is stable, i.e. $\operatorname{real}\left(
\xi\right)  =\xi^{\prime}<0$ for any $\xi\in\mathcal{S}$, provided that both
$M$ and $G$ are real, $M$ is antisymmetric and all diagonal elements of $G$
are positive.

The matrix $G$ can be decomposed as $G=\mathcal{A}+\mathcal{T}+\mathcal{D}$,
where $\mathcal{A}$ is antisymmetric, $\mathcal{T}$ is upper triangular, and
$\mathcal{D}$ is diagonal. The following holds $\mathcal{A}_{n,m}=G_{n,m}$ and
$\mathcal{T}_{n,m}=0$ for $n>m$, $\mathcal{A}_{n,m}=-G_{m,n}$ and
$\mathcal{T}_{n,m}=G_{n,m}+G_{m,n}$ for $n<m$, and for the diagonal elements
$\mathcal{A}_{n,n}=\mathcal{T}_{n,n}=0$ and $\mathcal{D}_{n,n}=G_{n,n}$. Using
this notation one has $M-G=\mathcal{M}-\mathcal{P}$, where $\mathcal{M}%
=M-\mathcal{A}$ and $\mathcal{P}=\mathcal{T}+\mathcal{D}$. As was shown above,
$M$ is antisymmetric provided that the Hamiltonian $\mathcal{H}$ is Hermitian,
hence $\mathcal{M}$ is antisymmetric as well. The eigenvalues of the upper
triangular matrix $\mathcal{P}$ are the diagonal elements of $G$, hence
$\mathcal{P}$ is positive definite provided that all diagonal elements of the
matrix $G$ are positive.

The real and imaginary parts of Eq. (\ref{v kappa}) are given by $\left(
\mathcal{M}-\mathcal{P}\right)  v^{\prime}=\xi^{\prime}v^{\prime}-\xi
^{\prime\prime}v^{\prime\prime}$ and $\left(  \mathcal{M}-\mathcal{P}\right)
v^{\prime\prime}=\xi^{\prime}v^{\prime\prime}+\xi^{\prime\prime}v^{\prime}$,
respectively, hence the following holds $\left(  v^{\prime}\right)
^{\mathrm{T}}\left(  \mathcal{M}-\mathcal{P}\right)  v^{\prime}+\left(
v^{\prime\prime}\right)  ^{\mathrm{T}}\left(  \mathcal{M}-\mathcal{P}\right)
v^{\prime\prime}=\xi^{\prime}\left(  \left(  v^{\prime}\right)  ^{\mathrm{T}%
}v^{\prime}+\left(  v^{\prime\prime}\right)  ^{\mathrm{T}}v^{\prime\prime
}\right)  $, or (recall that $\mathcal{M}$ is antisymmetric)%
\begin{align}
\xi^{\prime}  &  =\frac{\left(  v^{\prime}\right)  ^{\mathrm{T}}\left(
\mathcal{M}-\mathcal{P}\right)  v^{\prime}+\left(  v^{\prime\prime}\right)
^{\mathrm{T}}\left(  \mathcal{M}-\mathcal{P}\right)  v^{\prime\prime}}{\left(
v^{\prime}\right)  ^{\mathrm{T}}v^{\prime}+\left(  v^{\prime\prime}\right)
^{\mathrm{T}}v^{\prime\prime}}\nonumber\\
&  =-\frac{\left(  v^{\prime}\right)  ^{\mathrm{T}}\mathcal{P}v^{\prime
}+\left(  v^{\prime\prime}\right)  ^{\mathrm{T}}\mathcal{P}v^{\prime\prime}%
}{\left(  v^{\prime}\right)  ^{\mathrm{T}}v^{\prime}+\left(  v^{\prime\prime
}\right)  ^{\mathrm{T}}v^{\prime\prime}}\;,\nonumber\\
&
\end{align}
thus%
\begin{equation}
\xi^{\prime}\leq-\min G_{n,n}<0\;, \label{kappa' <=}%
\end{equation}
hence the system is stable.

\bibliographystyle{ieeepes}
\bibliography{acompat,Eyal_Bib}

\end{document}